\documentclass[12pt]{report}

\usepackage{multicol} 
\usepackage{latexsym}
\usepackage[svgnames]{xcolor} 

\usepackage{times} 

\usepackage{graphicx} 
\graphicspath{{figures/}} 
\usepackage{booktabs} 
\usepackage[font=small,labelfont=bf]{caption} 
\usepackage{amsfonts, amsmath, amsthm, amssymb} 
\usepackage{wrapfig} 
\title{ Nonlinear Cauchy-Riemann Equations and Liouville Equation For Conformal Metrics}

\author{Tu\u{g}\c{c}e Parlakg\"{o}r\"{u}r and Oktay K. Pashaev \\ \\
Department of Mathematics \\
Izmir Institute of Technology \\
 Izmir, 35430, Turkey}

\begin{document}

\maketitle

\begin{abstract}

 We introduce the Nonlinear Cauchy-Riemann equations as
 B\"{a}cklund transformations for
 several nonlinear and linear partial differential equations.
 From these equations we treat in details the Laplace and the Liouville equations
 by deriving general solution for the nonlinear Liouville equation.
 By M\"{o}bius transformation we relate solutions for the Poincare model of hyperbolic geometry, the Klein model in half-plane and
 the pseudo-sphere. Conformal form of the constant curvature metrics in these geometries, stereographic projections  and special solutions are discussed.
 Then we introduce the hyperbolic analog of the Riemann sphere, which we call the Riemann pseudosphere.
 We identify point at infinity on this pseudosphere and show that it can be used in complex analysis as an alternative to usual Riemann sphere to extend the complex plane.
 Interpretation of symmetric and antipodal points on both, the Riemann sphere and the Riemann pseudo-sphere, are
 given. By M\"{o}bius transformation and homogenous coordinates, the most general solution of Liouville equation as
 discussed by Crowdy is derived.

\end{abstract}

~
\newtheorem{thm}{Theorem}[subsection]
\newtheorem{cor}[thm]{Corollary}
\newtheorem{lem}[thm]{Lemma}
\newtheorem{prop}[thm]{Proposition}
\newtheorem{defn}[thm]{Definition}
\newtheorem{rem}[thm]{Remark}

\tableofcontents

\chapter{Introduction}

\section{Cauchy-Riemann Equations}

It is well know from complex analysis \cite{A}, if $u(x,y)$ and $v(x,y)$ are solutions of the Cauchy-Riemann (CR) equations
\begin{equation}
u_x = v_y,\,\,\,\,\,\,\,u_y = -v_x,\label{CR}
\end{equation}
in some domain, then both of them are harmonic functions
\begin{equation}
\Delta u = 0,\,\,\,\,\,\,\,\Delta v = 0.\label{LE}
\end{equation}
The first order system (\ref{CR}) plays the role of  the
B\"{a}cklund transformation for the second order Laplace
equation(\ref{LE}), since it relates two solutions of the Laplace
equation. This allows one from given solution of the Laplace
equation as a harmonic function,  construct another solution as a
harmonic conjugate one.

Combining two real functions as one complex function
$w(z)=u(x,y)+iv(x,y)$ we get analytic function, satisfying $\bar
\partial$ - equation
\begin{eqnarray}
\frac{\partial }{\partial \bar z} \ w= 0.\label{dbar0}
\end{eqnarray}
Then function $w(z)$ satisfies complex Laplace equation
\begin{eqnarray}
\Delta w = 0.\label{claplace}
\end{eqnarray}

\section{B\"{a}cklund Transformations}
The system of first order PDE relating two functions and often
depending of an additional parameter is called the B\"{a}cklund
Transformation. It implies that these two functions separately
satisfy the second-order PDE and each of them is then said to be a B\"{a}cklund
transformation of the other.

As an example, the Cauchy
Riemann Equations (\ref{CR}) are B\"{a}cklund Transformations for
the Laplace Equation. Let $ u(x,y)= x^2- y^2 $ is solution of
Laplace Equation (harmonic function), then by integrating CR equations we get another
(harmonically conjugate) solution of the Laplace equation $ v(x,y)= 2xy$.

\chapter{Nonlinear Cauchy-Riemann Equations}
The Cauchy-Riemann equations are linear equations and they allow to solve only linear Laplace equation.
Here we are going to generalize CR equations in such a way to be able solve nonlinear Laplace equations.

Let $u(x,y)$ and $v(x,y)$ are solutions of a system, which we call the Nonlinear Cauchy-Riemann (NCR) equations
\begin{eqnarray}
u_x = v_y + g(u,v),\,\,\,u_y = -v_x + f(u,v),\nonumber
\end{eqnarray}
where $f(u,v)$ and $g(u,v)$ are given functions.
According to the next theorem, these equations are B\"{a}cklund Transformations for Nonlinear Laplace Equations.

\section{Main Theorem}

\textbf{\textbf{Theorem 1. }} If real functions $u(x,y)$ and $v(x,y)$ satisfy the NCR equations
\begin{eqnarray}
u_x &=& v_y + g(u,v) \label{ncr},\\
u_y &=&-v_x + f(u,v)\label{ncr2},
\end{eqnarray}
where $g(u,v)$ and $f(u,v)$ are solutions of Cauchy-Riemann equations
\begin{eqnarray}
f_u = g_v,\,\,\,\,\
f_v =-g_u \label{CR f,g},
\end{eqnarray}
then $u(x,y)$ and $v(x,y)$ satisfy Nonlinear Laplace Equations
\begin{eqnarray}
\Delta u &=& \frac{1}{2}\frac{\partial}{\partial u}(f^2+g^2) \label{uv1},\\
\Delta v &=& \frac{1}{2}\frac{\partial}{\partial v}(f^2+g^2).\label{uv2}
\end{eqnarray}

\textbf{Proof}: By differentiating (\ref{ncr}),(\ref{ncr2}) in $x,y$, then adding and subtracting them we get
\begin{eqnarray}
u_{xx}+u_{yy}&=& g_{x}+f_{y},\nonumber\\
v_{xx}+v_{yy} &=&f_{x}-g_{y} \nonumber.
\end{eqnarray}
Applying Chain Rule to the r.h.s, where $f(u(x,y), v(x,y))$ and $g(u(x,y), v(x,y))$, we get
\begin{eqnarray}
u_{xx}+u_{yy}&=& (g_{u}u_x+g_{v}v_x)+(f_uu_y+f_vv_y),\nonumber\\
v_{xx}+v_{yy} &=&(f_uu_x+f_vv_x)-(g_uu_y+g_vv_y) \nonumber.
\end{eqnarray}
Substituting $u_x,u_y$ from (\ref{ncr}),(\ref{ncr2}) we get simply
\begin{eqnarray}
u_{xx}+u_{yy}&=& ff_{u}+gg_{u}\nonumber,\\
v_{xx}+v_{yy} &=&ff_{v}+gg_{v}.\nonumber
\end{eqnarray}
Finally we find
\begin{eqnarray}
\Delta u &=& \frac{1}{2}\frac{\partial}{\partial u}(f^2+g^2)\nonumber ,\\
\Delta v &=& \frac{1}{2}\frac{\partial}{\partial v}(f^2+g^2)\nonumber.\\ &\, & \hskip5cm\,\,\,\,\,\,\,\,\,\,\,\,\,\,\,\,\,\Box \nonumber
\end{eqnarray}

This B\"{a}cklund transformations and Nonlinear Laplace equations acquire quite simple form in terms of complex functions.

\textbf{\textbf{Corollary 1. }} Let
\begin{eqnarray}
w(x,y)=u(x,y)+iv(x,y)\nonumber
\end{eqnarray}
is complex valued function of $(x,y)$
and
\begin{eqnarray}
G(w)=f(u,v)+ig(u,v)\nonumber
\end{eqnarray}
is complex valued function of $(u,v)$.
Then due to the Cauchy-Riemann equations (\ref{CR f,g}), function
$G(w)$ is analytic, so that
\begin{eqnarray}
\frac{\partial}{\partial \bar w} \ G(w)=0,\nonumber
\end{eqnarray}
where $\frac{\partial}{\partial \bar w}=\frac{1}{2}(\frac{\partial}{\partial u}+i\frac{\partial}{\partial v})$.

\subsection{Complex Form}

Due to this corollary, every analytic function $G(w)$ determines NCR equations (\ref{ncr}), (\ref{ncr2}), where $f(u,v) = ${Re}$\, G(w) $, $g(u,v) = ${Im}$\, G(w)$.
These equations can be rewritten as one complex $\bar \partial$ - equation
\begin{eqnarray}
\frac{\partial }{\partial \bar z} \ w=\frac{i}{2} \,\,\overline{G(w)},\label{dbar}
\end{eqnarray}
  and the system (\ref{uv1}), (\ref{uv2}) as one complex Nonlinear Laplace Equation
\begin{eqnarray}
\triangle w =\overline{G'(w)}\, G(w).\label{nlaplace}
\end{eqnarray}
Real and imaginary parts of this equation give us Nonlinear Laplace Equations for $u(x,y)$ and $v(x,y)$, related by B\"{a}cklund Transformation  as NCR.

 Below we illustrate these results by several examples.

\textbf{Example 1.}
As a first example we consider the trivial case $ f=0, g=0$ and as follows $G=0$. In this case we have the linear CR equations relating solutions of linear Laplace equation, which we have already discussed.

\subsection{Helmholz Equation}

\textbf{Example 2.}
The choice $G(w)=w$, implies the linear, but modified CR type equations
\begin{eqnarray}
u_x &=& v_y + v,\nonumber\\
u_y &=& -v_x + u,\nonumber
\end{eqnarray}
 connecting solutions of the linear Helmholz Equation $$\Delta u=u,\,\,\,\,\,\Delta v=v .$$ By solving the first order system we can construct new solution of the Helmholz Equation from the given one. The above NCR equations we can rewrite as one complex  $\bar \partial$- equation $$\frac{\partial
w}{\partial \bar z}=\frac{i}{2}\bar w.$$
It implies the complex nonlinear Laplace equation $\Delta w = w$.

\subsection{Nonlinear Laplace Equations}

\textbf{Example 3.}
If we choose $G(w)=w^2$, then we get NCR equations
\begin{eqnarray}
u_x &=& v_y + 2uv,\nonumber\\
u_y &=& -v_x + u^2-v^2,\nonumber
\end{eqnarray}
or one complex  $\bar \partial$- equation $$\frac{\partial
w}{\partial \bar z}=\frac{i}{2}\bar w^2.$$

 It implies the complex cubic nonlinear Laplace Equation (the nonlinear Schrödinger equation)
  $$\Delta w
=2|w|^{2}w$$
equivalent to the coupled system of  nonlinear Laplace equations
with cubic nonlinearity
\begin{eqnarray}
 \Delta u&=&2(u^2+v^2)u,\nonumber \\
 \Delta v&=&2(u^2+v^2)v.\nonumber
\end{eqnarray}

\textbf{Example 4.}
More general function $G(w)=w^n$ gives NCR equations
\begin{eqnarray}
u_x &=& v_y + {Im}\, (u+iv)^n,\nonumber\\
u_y &=& -v_x + {Re}\, (u+iv)^n,\nonumber
\end{eqnarray}
or in complex form, the $\bar \partial$ - equation $$\frac{\partial
w}{\partial \bar z}=\frac{i}{2}\bar w^n.$$

It implies complex nonlinear Laplace Equation with higher nonlinearity
$$\triangle w
=n|w|^{2(n-1)}w,$$ which is equivalent to the coupled system of nonlinear
Laplace equations
\begin{eqnarray}
\Delta u=n\,(u^2+v^2)^{n-1}u, \nonumber\\
\Delta v=n\,(u^2+v^2)^{n-1}v.\nonumber
\end{eqnarray}

\chapter{Liouville Equation}

\section{The Liouville Equation}
If we choose $G(w)= e^{w} $, then we get NCR
\begin{eqnarray}
u_x &=& v_y + e^u \sin v, \label{BT Liouville2} \\
u_y &=& -v_x +  e^u \cos v, \label{BT Liouville}
\end{eqnarray}
or in complex form, the $\bar \partial$ - equation
\begin{equation}\frac{\partial w}{\partial \bar z}=\frac{i}{2}e^{\bar w}.
\label{LDBAR}\end{equation} It is a B\"{a}cklund transformation
relating linear Laplace equation with the nonlinear Liouville
equation \cite{L}. First we get complex Nonlinear Laplace Equation
$$\Delta w =e^{w+ \bar w}$$ and then, separating real and imaginary
parts,  we have decoupled system of the Liouville equation and the
linear Laplace equation correspondingly
\begin{eqnarray}
\Delta u&=&e^{2u},\nonumber \\
\Delta v&=&0.\nonumber
\end{eqnarray}

\section{General Solution of Liouville Equation}
By using NCR (\ref{BT Liouville2}),(\ref{BT
Liouville}) we can construct general solution of Liouville equation
in terms of general solution of the Laplace equation.

\textbf{Theorem 2.}
 The general solution of Liouville equation
\begin{eqnarray}
\Delta u&=&e^{2u}\nonumber
\end{eqnarray}
is
\begin{eqnarray}
u(x,y)=\frac{1}{2}\ln\frac{4\frac{d A}{d z}\frac{d \bar A}{d \bar z}}{(A(z)+ \bar A(\bar z))^2},\label{generalliouville}
\end{eqnarray}
where $A(z)$ is an arbitrary analytic function of $z = x + i y$.

\textbf{Proof}: the general solution of Laplace equation
$$v(x,y)=F(z)+\overline {F(z)}$$ is determined in terms of an
arbitrary analytic function $F(z)$. Substituting this solution to
equation (conjugate of $\bar \partial$ equation)
\begin{eqnarray}
\frac{\partial \bar w}{\partial  z}= -\frac{i}{2}e^{w},\nonumber
\end{eqnarray}
we get
\begin{eqnarray}
\frac{\partial u}{\partial z}-i\frac{\partial F(z)}{\partial
z}=-\frac{i}{2}e^u e^{i(F(z)+\overline{F(z)})}.\nonumber
\end{eqnarray}
Let us define $\phi = e^u$ so that
\begin{eqnarray}
\frac{\partial \phi}{\partial z}-i\phi \frac{\partial F(z)}{\partial
z}=-\frac{i}{2}\phi^2 e^{i(F(z)+\overline{F(z)})}.\nonumber
\end{eqnarray}
Introducing new function $\psi$, according to formula $\phi=
\psi e^{i(F(z)+\overline{F(z)})}$, we have
equation
\begin{eqnarray}
\frac{\partial \psi}{\partial z}=-\frac{i}{2}\psi^2
e^{2i(F(z)+\overline{F(z)})}.\nonumber
\end{eqnarray}
By multiplying both sides with $-1/\psi^2$ we have
\begin{eqnarray}
\frac{\partial }{\partial z}\frac{1}{\psi}=\frac{i}{2}
e^{2i(F(z)+\overline{F(z)})}.\nonumber
\end{eqnarray}
and after integrating we get
\begin{eqnarray}
\frac{1}{\psi}=\frac{i}{2} e^{2i\overline{F(z)}}\left[\int^z
e^{2iF(z')}dz'+G(\bar z)\right].\nonumber
\end{eqnarray}
For function $\phi$ it gives
\begin{eqnarray}
\phi=\frac{e^{i(F(z)-\overline{F(z)})}}{\frac{i}{2} \int^z
e^{2iF(z')}dz'+\frac{i}{2}G(\bar z)}.\nonumber
\end{eqnarray}
Since function $F(z)$ is an arbitrary analytic function, instead of
it we can introduce another analytic function
\begin{eqnarray}
A(z)={\frac{i}{2} \int^z e^{2iF(z')}dz'}\label{az}
\end{eqnarray}
and instead of arbitrary $G(\bar z)$ a new function
\begin{eqnarray}
\overline {B(z)}= \frac{i}{2}G(\bar z).\nonumber
\end{eqnarray}
Then our solution becomes
\begin{eqnarray}
\phi=\frac{e^{i(F(z)-\overline{F(z)})}}{A(z)+\overline{B(z)}}.\nonumber
\end{eqnarray}
Due to reality condition: $\phi=\bar \phi$ we find $A(z)=B(z)$. More general solution $A(z)=B(z)+c $, where $c$ is an arbitrary real constant, can be reduced also to this case.
By differentiating (\ref{az}) we get
\begin{eqnarray}
e^{iF(z)}=\left(\frac{2}{i}\frac{d A}{d
z}\right)^{1/2}.\nonumber
\end{eqnarray}
Then, we obtain solution  in terms of $A(z)$ as
\begin{eqnarray}
\phi=2\frac{\left(\frac{d A}{d z}\frac{d \bar
A}{d \bar z}\right)^{1/2}}{A(z)+\overline{A(z)}}.\nonumber
\end{eqnarray}
Finally, we find the general solution of Liouville equation
\begin{eqnarray}
\Delta u&=&e^{2u}\nonumber
\end{eqnarray}
in the form
\begin{eqnarray}
u(x,y)=\frac{1}{2}\ln\frac{4\frac{d A}{d z}\frac{d
\bar A}{d \bar z}}{(A(z)+ \bar A(\bar z))^2}.\label{*}
\end{eqnarray}
By direct substitution we check that it is really solution of the
Liouville equation, determined in terms of an arbitrary analytic
function $A(z)$. Now by choosing function $A(z)$ in a different way we
can construct several particular solutions of Liouville equation.

\textbf{Example 1.} Let function $A$ has a simple zero at
origin $A(z)=z$. Substituting into equation (\ref{generalliouville})
 we get particular solution of Liouville equation as $u(x, y)=\frac{1}{2}\ln(\frac{1}{x^2}).$ \\

\textbf{Example 2.} By choosing function $A(z)=\frac{1}{z}$, with
simple pole at origin, we get the same result
$u(x,y)=\frac{1}{2}\ln(\frac{1}{x^2}).$  \\

\textbf{Example 3.}
For double zero  $A(z)=z^2$ and double pole $A(z)=\frac{1}{z^2}$ cases, we find the same solution of Liouville equation as
$$u(x,y)=\frac{1}{2}\ln\frac{4(x^2+y^2)}{(x^2-y^2)^2}.$$
The solution is singular for $x=\pm y$. The plot of this solution is
given in Figure 3.1\\

\begin{center}\vspace{0.25cm}
\includegraphics[width=0.80\linewidth]{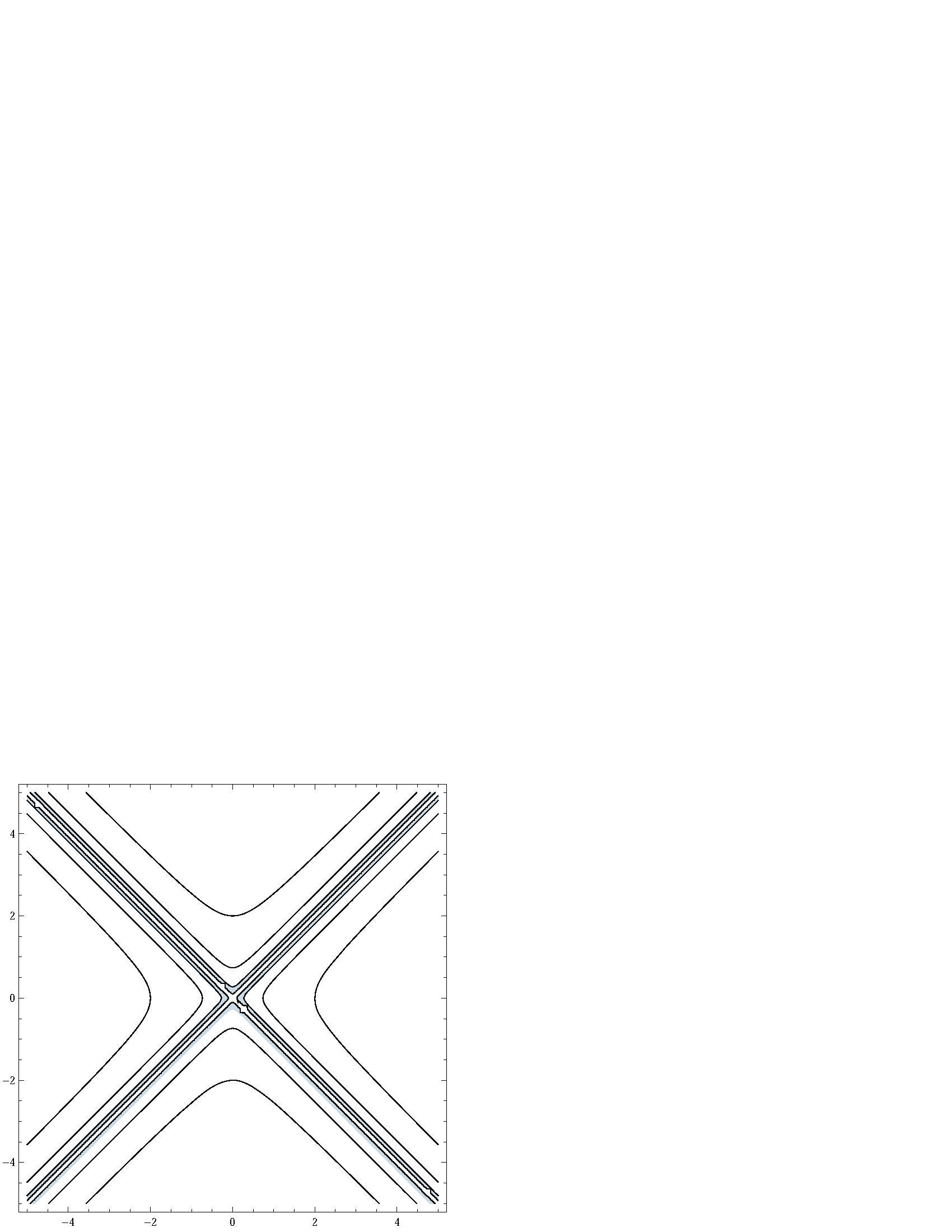}
\captionof{figure}{\color{Navy}Liouville solution for $z^2$}
\end{center}\vspace{0.25cm}

\textbf{Example 4.} Generalizing to higher order
zero  $A(z)=z^n$ and higher order pole $A(z)=\frac{1}{z^n}$ we get
result
\begin{eqnarray}
u=\frac{1}{2}\ln\frac{4n^2|z|^{2n-2}}{(z^n+ (\bar z)^n)^2}.\nonumber
\end{eqnarray} \\
In polar coordinates $z=r e^{i\theta}$, $r> 0, $ $0 \le \theta \le 2 \pi$, this solution becomes
\begin{eqnarray}
u=\frac{1}{2}\ln\frac{n^2}{r^2 \cos^2 n\theta }.\nonumber
\end{eqnarray} \\

For $n=2$ in polar form we have
\begin{eqnarray}
u=\frac{1}{2}\ln\frac{2}{r^2 \cos^2 2\theta }.\nonumber
\end{eqnarray} \\

This solution is singular for $r=0$ and for lines: $r > 0$, $\cos 2\theta=0$ or $\theta=\pi
/4+\pi k/2 $, $k = 0, 1, 2, 3$. It corresponds to two lines $x = \pm y$.
The plot of this solution is given in
Figure3.1

For $n=3$ we have
\begin{eqnarray}
u=\frac{1}{2}\ln\frac{9(x^2+y^2)^2}{x^2(x^2-3y^2)^2 }.\nonumber
\end{eqnarray} \\
This solution is singular for $x=\pm \sqrt{3}y$ and $x=0$.
Also in polar form we have
\begin{eqnarray}
u=\frac{1}{2}\ln\frac{2}{r^2 \cos^2 3\theta },\nonumber
\end{eqnarray} \\
which is singular for lines: $r > 0$, $\cos 3\theta=0$ or $\theta=\pi
/6+\pi k/3 $, $k = 0, ..., 5$. This corresponds to three lines: $x=\pm \sqrt{3}y$ and $x=0$.
The plot of this solution is given in
Figure 3.2

\begin{center}\vspace{0.25cm}
\includegraphics[width=0.80\linewidth]{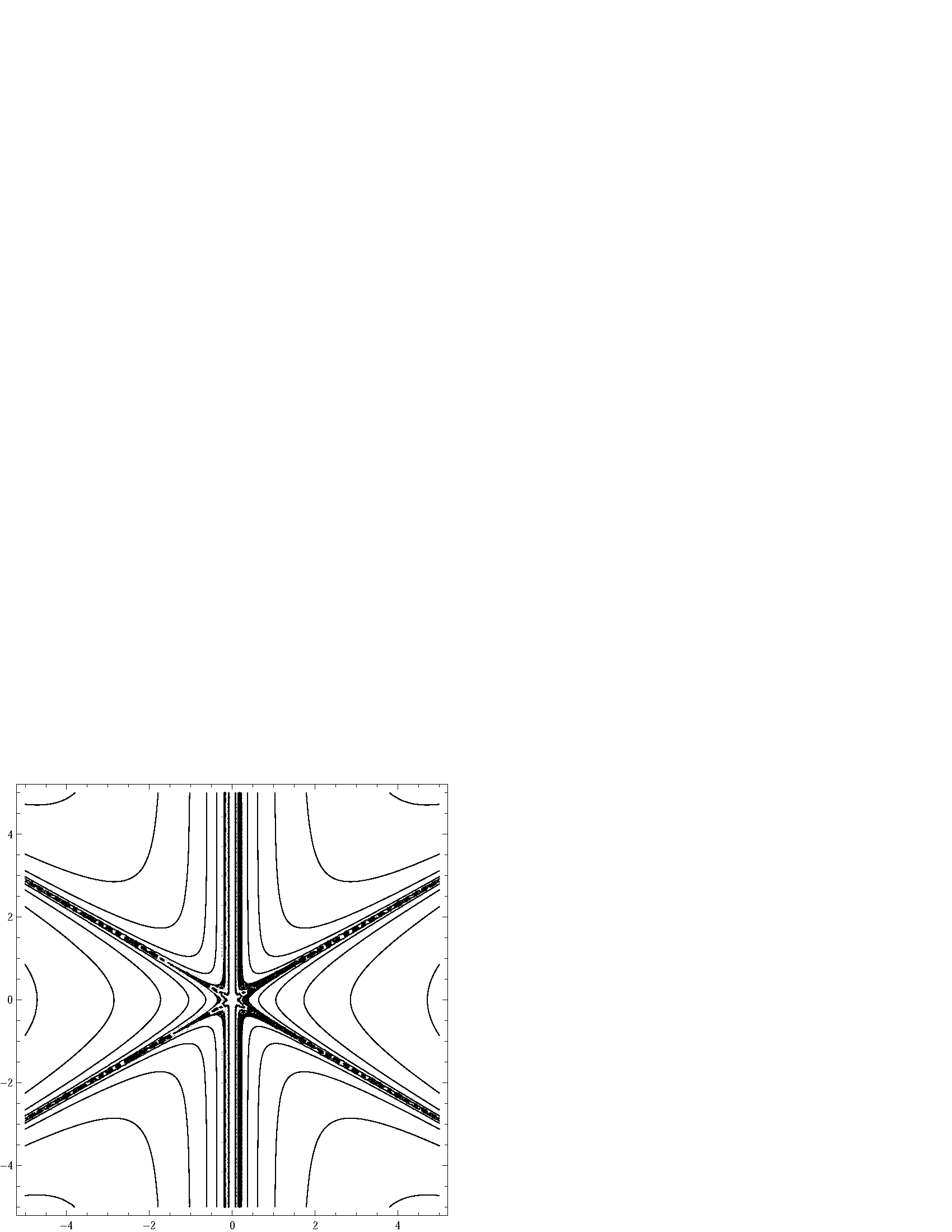}
\captionof{figure}{\color{Navy}Liouville solution for $z^3$}
\end{center}\vspace{0.25cm}

For $n$- an arbitrary positive integer, we have singular lines at angles: $\cos n \theta = 0$,
or $\theta=\frac{\pi}{2 n} + \frac{\pi}{n} k $, $k = 0, ..., 2n-1$. This corresponds to $n$-lines
in $xy$-plane, passing thought the origin, and with angle $\frac{\pi}{n}$, fixed between lines.\\

\textbf{Example 5.}
If $A(z)=e^z$, then it has no zeros or poles. But it is periodic due to $e^{z + 2\pi i} = e^z$. Then, we find periodic solution as $u=-\ln(\cos y)$. \\

\begin{center}\vspace{0.25cm}
\includegraphics[width=0.40\linewidth]{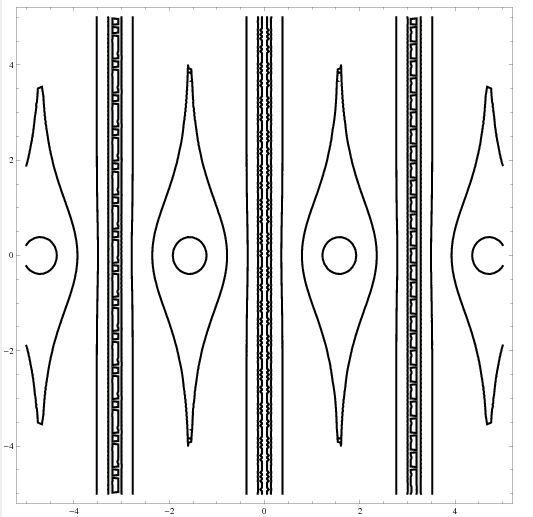}
\captionof{figure}{\color{Navy}Contour Plot of periodic solution}
\end{center}\vspace{0.25cm}

\textbf{Example 6.} For $A(z)=\sin z$, solution is
\begin{eqnarray}
u=\frac{1}{2}\ln \frac{4\cos z \cos \bar z}{(\sin z+ \sin \bar z)^2}\nonumber
\end{eqnarray}

or in real variables

\begin{center}\vspace{0.25cm}
\includegraphics[width=0.50\linewidth]{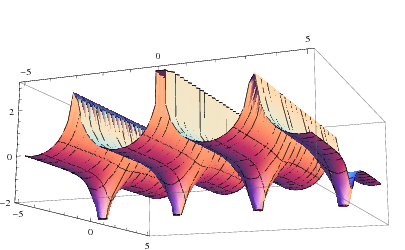}
\captionof{figure}{\color{Navy}3D Plot of periodic solution}
\end{center}\vspace{0.25cm}

\begin{eqnarray}
u=\frac{1}{2}\ln (\cot^2 x + \tanh^2 y).\nonumber
\end{eqnarray}
It is periodic in $x$ variable (see Figures 3.3 and 3.4) \\

\subsection{Poles and Zeros Symmetry of Solution }\label{221}
The general solution
\begin{eqnarray}
u=\frac{1}{2}\ln\frac{4\frac{d A}{d z}\frac{d
\bar A}{d \bar z}}{(A(z)+ \bar A(\bar z))^2}\nonumber
\end{eqnarray}
under substitution $$A(z)= \frac{1}{H(z)}$$ becomes of the same form
\begin{eqnarray}
u=\frac{1}{2}\ln\frac{4\frac{d H}{d z}\frac{d
\bar H}{d \bar z}}{(H(z)+ \bar H(\bar z))^2}. \nonumber
\end{eqnarray}
It implies that zeros and poles of function $A(z)$ give the same
solution. For example $A(z)=z^n$ and $A(z)=1/z^n$ give the same $u$.
According to this, it is sufficient to look only the case of zeros.

\subsection{{Meromorphic Solution} }
If  $$A(z)= \frac{C(z)}{D(z)}, $$ (for polynomials $C(z)$ and $D(z)$ it is a meromorphic
function),  we have solution
\begin{eqnarray}
u=\frac{1}{2}\ln\frac{4|C'D-CD'|^2}{(C \overline{ D} + \overline{ C}
D)^2}.\nonumber
\end{eqnarray}
As easy to see, this formula is symmetric under replacement $C$
$\leftrightsquigarrow$ $D$.

Next we construct several explicit solutions.

\textbf{Example 1.} Let choose $A(z)= \tan z$ and $A(z)= \cot z$ , giving the same meromorphic solution

\textbf{Example 2.} Choosing $A(z) = \csc z$, then $H(z)= \sin z$. Clearly, zeros of $H(z)$ becomes poles of $A(z)$.

\textbf{Example 2.} Choosing $A(z) = \sec z$, then $H(z)= \cos z$. Clearly, zeros of $H(z)$ becomes poles of $A(z)$.

\subsection{Liouville Equations with Complex Coefficients }
Instead of $G(w)=e^w$ we can consider slightly more general choice, depending
on complex coefficients.

 \textbf{Case 1.} Let $G(w)=be^w$, where $b$ is a
complex number. Then, calculating $\overline{G'(w)}$ and $G(w)$ we
find $\Delta w=|b|^2e^{\bar w +w}$. Thus $\Delta u=|b|^2e^{2u}$ and
$\Delta v=0$. Equation for $u$ is in a more general form of the Liouville
equation and as easy to see it can be reduced to the
Liouville equation $\Delta' u=e^{2u}$ by simple
coordinate changes $|b| x=x'$, $|b| y=y'$. \\

\textbf{Case 2.} Let $G(w)=e^{aw}$, where $a$ is a complex number,
say $a=m+in$. So $\Delta w=\bar a e^{\bar a \bar w +aw}$
$\Rightarrow$ $\Delta (u+iv)=\bar a e^{u(\bar a+a)+iv(a-\bar a)}$.
For the real and imaginary parts it gives

$$\Delta u=me^{2( mu-nv)} ,\,\,\,\,\Delta v=-ne^{2( mu-nv)}.$$
 Let
$\mu = (mu-nv) $, then multiplying equations by $n$ and $m$
respectively, and subtracting, we get
\begin{eqnarray}
\Delta \mu = (m^2+n^2) e^{2\mu}.\nonumber
\end{eqnarray}
Since $a=m+in$ , $|a|^2= m^2+n^2$ and then  we have $\Delta \mu = |a|^2
e^{2\mu}$. Also for $\Delta v=-ne^{2( mu-nv)}$ applying the same
procedure we get
\begin{eqnarray}
\Delta (nu+mv)= 0.\nonumber
\end{eqnarray}
Denoting $\nu= (nu+mv)$ we find $\Delta \nu = 0$. Finally we obtain
the system as in Example 1:
\begin{eqnarray}
\Delta \mu &=& |a|^2 e^{2\mu},\label{delta Mu} \\
\Delta \nu &=& 0.\label{delta Nu}
\end{eqnarray}

Thus,by substitution $|a| x=x'$, $|a| y=y'$ the system  (\ref{delta Mu}) , (\ref{delta Nu}) can be reduced to the
Liouville equation $\Delta' \mu=e^{2\mu}$ and the Laplace equation
$\Delta' \nu= 0$. \\

\textbf{Case 3.} Choosing $G(w)=be^{aw}$, where $a=m+in$ and $b$
are an arbitrary complex numbers,
we have $\Delta w=\bar a|b|^2e^{\bar a \bar w +aw}$ $\Rightarrow$\\

$\Delta (u+iv)=\bar a|b|^2 e^{u(\bar a+a)+iv(a-\bar a)}$
$\Rightarrow$ $\Delta (u+iv)=(m-in)|b|^2 e^{2(mu-nv)}.\\$ For real
and imaginary parts we have

$$\Delta u=m|b|^2 e^{2( mu-nv)},\,\,\,\,\,\,\,\Delta v=-n|b|^2 e^{2(
mu-nv)}.$$
 Introducing as before $\mu =(mu-nv)$,
multiplying equations by $m/|b|^2$ and $n/|b|^2$ respectively, and
subtracting we find
\begin{eqnarray}
\Delta \mu = (m^2+n^2)|b|^2 e^{2\mu}.\nonumber
\end{eqnarray}
Since $a=m+in$ , $|a|^2= m^2+n^2$ and we have $\Delta \mu = |a|^2 |b|^2 e^{2\mu}$.\\

 Applying the same
procedure to equation $\Delta v=-n|b|^2 e^{2( mu-nv)}$,
\begin{eqnarray}
\Delta (nu+mv)= 0\nonumber
\end{eqnarray}
for $\nu= (nu+mv)$ we get $\Delta \nu = 0\,.$

So we find the system
\begin{eqnarray}
\Delta \mu &=& |a|^2 |b|^2 e^{2\mu},\label{delta Mu2} \\
\Delta \nu &=& 0\,.\label{delta Nu2}
\end{eqnarray}

The system (\ref{delta Mu2}) , (\ref{delta Nu2}) can be reduced to
the Liouville equation $\Delta' \mu=e^{2\mu}$ and the Laplace
equation $\Delta' \nu= 0$ by substitution
$|a||b| x=x'$, $|a||b| y=y'$. \\ \\

\section{B\"{a}cklund Transformation Depending on Parameter}

The above consideration allow us to extend the  B\"{a}cklund
transformation (\ref{BT Liouville}), (\ref{BT Liouville2}), to
include an arbitrary real parameter. We choose $G(w) = e^{i \lambda}
e^{w}$, where $\lambda$ is real, and which corresponds to the \textbf{Case 1.},
with $b = e^{i \lambda}$. Then we get the B\"{a}cklund
transformation
\begin{eqnarray}
u_x &=& v_y + e^u \sin (v + \lambda), \label{LBT Liouville2} \\
u_y &=& -v_x +  e^u \cos (v + \lambda), \label{LBT Liouville}
\end{eqnarray}
and the $\bar \partial$ - equation
\begin{equation}\frac{\partial w}{\partial \bar z}=\frac{i}{2}e^{-i\lambda} e^{\bar w}
\label{LLDBAR}\end{equation} depending on parameter $\lambda$. As
easy to see, the Laplace and Liouville equations are independent of
this parameter.
\section{Conformal Transformation of Liouville Equation}
Here we show that the Liouville equation is invariant under conformal transformation
generated by an arbitrary analytic function $w = w(z)$.
\subsection{Conformal Transformation of Laplace Equation}
We start with Laplace equation for an arbitrary harmonic function
$v(x,y)$:  $\Delta v= 0$. Let $w=w(z)$ (or $z=z(w)$) is analytic
function satisfying
\begin{eqnarray}
\frac{\partial }{\partial \bar z}w = 0,\nonumber \\
\frac{\partial }{\partial \bar w}z = 0.\nonumber
\end{eqnarray}
In terms of these variables our function becomes $v(z(w),\overline
{z(w)})=V(w,\bar w)$. Since $\Delta v(z,\bar z)= 0$ we will change the
function into $V$ by using Chain rule
\begin{eqnarray}
\frac{\partial}{\partial z} = \frac{dw}{dz} \frac{\partial}{\partial
w},\nonumber
\end{eqnarray}
\begin{eqnarray}
\frac{\partial}{\partial \bar z} = \frac{d \bar w}{d\bar z}
\frac{\partial}{\partial \bar w}.\nonumber
\end{eqnarray}
Substituting  these into Laplace Equation we get
\begin{eqnarray}
4 \frac{\partial}{\partial z} \frac{\partial}{\partial \bar z} v = 4
\left(\frac{dw}{dz} \frac{\partial}{\partial w} \right)
\left(\frac{d \bar w}{d\bar z} \frac{\partial}{\partial \bar
w}\right) V = 0\nonumber
\end{eqnarray}
\begin{eqnarray}
\left|\frac{dw}{dz}\right|^2 4 \frac{\partial}{\partial w}
\frac{\partial}{\partial \bar w} V = 0 ,\nonumber
\end{eqnarray}
clearly
\begin{eqnarray}
\Delta V(w, \bar w) = 0 .\nonumber
\end{eqnarray}

\subsection{Conformal Transformation of Liouville Equation}
The Liouville equation is $\Delta u= 4 \frac{\partial}{\partial z}\frac{\partial}{\partial \bar z} u = e^{2u}$, where $z=x+iy$. Since $w=w(z)$ (or $z=z(w)$) we have
\begin{eqnarray}
u(z(w),\overline {z(w)})= \phi(w,\bar w).\label {U}
\end{eqnarray}
Then, applying the same procedure as above we get
\begin{eqnarray}
\left|\frac{dw}{dz}\right|^2 4 \frac{\partial}{\partial w} \frac{\partial}{\partial \bar w} \phi = e^{2\phi}.\nonumber
\end{eqnarray}
Therefore
\begin{eqnarray}
\Delta \phi =\left|\frac{dz}{dw}\right|^2 e^{2\phi},\nonumber
\end{eqnarray}
and this is equal to
\begin{eqnarray}
\Delta \phi = e^{2\phi + \ln {\left|\frac{dz}{dw}\right|^2}}.\nonumber
\end{eqnarray}
Now let us define
\begin{eqnarray}
 U =\phi + \frac{1}{2}\ln {\left|\frac{dz}{dw}\right|^2}.\label {Phi}
\end{eqnarray}
Taking Laplacian from both sides we get the following expression
\begin{eqnarray}
 \Delta U = \Delta \phi + \Delta \left( \frac{1}{2}\ln {\left|\frac{dz}{dw}\right|^2}\right).\nonumber
\end{eqnarray}
It is easily seen that
\begin{eqnarray}
\Delta \left( \frac{1}{2}\ln {\left|\frac{dz}{dw}\right|^2}\right) = 0 .\nonumber
\end{eqnarray}
Clearly
\begin{eqnarray}
 \Delta U = \Delta \phi .\nonumber
\end{eqnarray}
Thus,
\begin{eqnarray}
\Delta U = e^{2U},\nonumber
\end{eqnarray}
where $w=u+iv$. From (\ref {U}) we get the result
\begin{eqnarray}
\Delta u = e^{2 u} \Rightarrow \Delta u= u_{xx}+u_{yy} = e^{2 u},\nonumber
\end{eqnarray}
\begin{eqnarray}
\Delta U = e^{2 U} \Rightarrow \Delta U = U_{uu}+ U_{vv} = e^{2 U},\nonumber
\end{eqnarray}
where $u$ and $U$ are related by
\begin{eqnarray}
U(w, \bar w) = u(z, \bar z) + \frac{1}{2} \ln {\left|\frac{dz}{dw}\right|^2}.\label{ph}
\end{eqnarray}
\textbf{Example .}
Let
\begin{eqnarray}
u(z, \bar z)=\frac{1}{2}\ln\frac{4}{(z + \bar z)^2}, \nonumber
\end{eqnarray}
and we choose $w = \ln z$ $\Rightarrow$ $z=e^w$. Substituting $z$ into $u$ and then using equation from (\ref{ph})
we get the result in terms of $(w, \bar w)$, such that
\begin{eqnarray}
U(w, \bar w)=\ln\frac{2}{(e^w + e^{\bar w})} + \frac{w + \bar w}{2}. \nonumber
\end{eqnarray}
It satisfies the Liouville equation $\Delta U = e^{2 U}$.

\section{General Solution of Liouville Equation Under Conformal Transformation}

For Liouville equation $\Delta u = e^{2 u}$ we have the general solution
\begin{eqnarray}
\psi(z, \bar z)=\frac{1}{2}\ln\frac{4\frac{d A}{d z}\frac{d
\bar A}{d \bar z}}{(A(z)+ \bar A(\bar z))^2}.\nonumber
\end{eqnarray}
Now suppose $w=w(z)$ is analytic and $u (z,\bar z) = u (z(w),\overline {z(w)})= \phi(w,\bar w)$.
Then
\begin{eqnarray}
A(z(w))\equiv B(w)\nonumber
\end{eqnarray}
\begin{eqnarray}
\overline {A(z(w))}\equiv \overline {B(w)}.\nonumber
\end{eqnarray}
Since $A$ transforms into $B$, we have
\begin{eqnarray}
\frac{d A}{d z} = \frac{d A}{d w}\frac{dw}{dz}  = \frac{d B}{d w}\frac{dw}{dz},\nonumber
\end{eqnarray}
and taking conjugate of this equation
\begin{eqnarray}
\frac{d \bar A}{d \bar z} = \frac{d \bar A}{d \bar w}\frac{d \bar w}{d \bar z}  = \frac{d \bar B}{d \bar w}\frac{d \bar w}{d \bar z}.\nonumber
\end{eqnarray}
Now we are ready to write solution of Liouville equation in terms of $w$ coordinates
\begin{eqnarray}
\phi(w, \bar w)=\frac{1}{2}\ln\frac{4\left|\frac{dw}{dz}\right|^2\frac{d B}{d w}\frac{d
\bar B}{d \bar w}}{(B(w)+ \bar B(\bar w))^2}.\nonumber
\end{eqnarray}
To simplify this equation we split
\begin{eqnarray}
\phi(w, \bar w)=\frac{1}{2}\ln\frac{4\frac{d B}{d w}\frac{d
\bar B}{d \bar w}}{(B(w)+ \bar B(\bar w))^2}+\frac{1}{2}\ln \left|\frac{dw}{dz}\right|^2.\label{Ps}
\end{eqnarray}
Since
\begin{eqnarray}
\frac{1}{2}\ln \left|\frac{dw}{dz}\right|^2 = \frac{1}{2}\ln \left|\left(\frac{dz}{dw}\right)^{-1}\right|^2,\nonumber
\end{eqnarray}
we have
\begin{eqnarray}
\frac{1}{2}\ln \left|\frac{dw}{dz}\right|^2 = -\frac{1}{2} \ln  \left|\frac{dz}{dw}\right|^2 .\nonumber
\end{eqnarray}
Substituting these into equation (\ref{Ps}) we get
\begin{eqnarray}
\phi(w, \bar w)+ \frac{1}{2} \ln  \left|\frac{dz}{dw}\right|^2 = \frac{1}{2}\ln\frac{4\frac{d B}{d w}\frac{d
\bar B}{d \bar w}}{(B(w)+ \bar B(\bar w))^2},\nonumber
\end{eqnarray}
then comparing with (\ref {Phi}) we find that function
\begin{eqnarray}
U= \frac{1}{2}\ln\frac{4\frac{d B}{d w}\frac{d
\bar B}{d \bar w}}{(B(w)+ \bar B(\bar w))^2}.\nonumber
\end{eqnarray}
is solution of Liouville equation $\Delta U = e^{2 U}$. Due to arbitrariness of $A(z)$ and as follows $B(w)$, this is the general solution of Liouville equation in new coordinates.

\section{Poincare and Klein Models}\label{231}
Here we consider particular case of conformal mapping as M\"obius transformation from
half plane to the unit disk. Transformation
\begin{eqnarray}
w=\frac{1+z}{1-z}\nonumber
\end{eqnarray}
maps the right half plane $z$ to the unit disk $|w|=1$. So, that symmetric points
$-1$ and $1$ under vertical axis transform to symmetrical points
$0$ and $\infty$ under unit circle $|w|=1$. This transformation
relates to different geometries: the Klein geometry in half plane
and the Poincare geometry in unit disk. The above transformation
implies substitution for solution of Liouville equation
\begin{eqnarray}
A(z)=\frac{1+B(z)}{1-B(z)},\label{mob}
\end{eqnarray}
which corresponds to the M\"obius transformation between complex plane $B$ and complex plane $A$. Under this
transformation, the general solution of Liouville equation (\ref{*}) becomes
\begin{eqnarray}
u=\frac{1}{2}\ln\frac{4\frac{d B}{d z}\frac{d
\bar B}{d \bar z}}{(1-B(z) \overline {B(z)})^2}.\label{**}
\end{eqnarray}
This map transforms the right half plane $Re A>0$ to the unit disk
$|B|<1$. It is shown in Figure 3.5.

\begin{center}\vspace{0.25cm}
\includegraphics[width=0.80\linewidth]{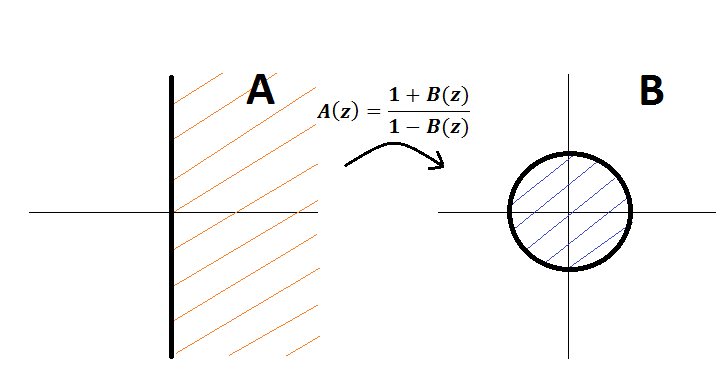}
\captionof{figure}{\color{Navy}Klein and Poincare Geometry}
\end{center}\vspace{0.25cm}

Now we will consider several examples by using (\ref{**}) .

\textbf{Example 1.} Let $B(z)=z$, then applying the above formula we
get
\begin{eqnarray}
u=\ln{\frac{2}{|1-|z|^2|}}.\nonumber
\end{eqnarray}
It is regular inside of open disk $|z|<1$, where $u=\ln{\frac{2}{1-|z|^2}}$. For exterior of unit disk $|z|>1$ it is $u=\ln{\frac{2}{|z|^2-1}}$. On the unit circle $|z|=1$ , solution becomes singular.\\
Writing the solution in $xy$ coordinates we get
\begin{eqnarray}
u=\ln{\frac{2}{|1-x^2-y^2|}}.\nonumber
\end{eqnarray}

\textbf{Example 2.} In more general case, let $B(z)=z^n$, where $n>0$ and we have
\begin{eqnarray}
u=\ln{\frac{2n |z|^{n-1}}{1-|z|^{2n}}}.\nonumber
\end{eqnarray}
Choosing $z$ in a polar form, such that $z=r e^{i\theta}$, then our solution becomes
\begin{eqnarray}
u=\ln{\frac{2n r^{n-1}}{1-r^{2n}}}.\nonumber
\end{eqnarray}

If we choose $n<0$ as $n\Rightarrow -n$ the solution is
\begin{eqnarray}
u=\ln{\frac{-2n |z|^{-n-1}}{1-|z|^{-2n}}}\nonumber
\end{eqnarray}
and after simplification
\begin{eqnarray}
u=\ln{\frac{2n |z|^{n-1}}{1-|z|^{2n}}}.\nonumber
\end{eqnarray}
It looks the same as solution whith $n>0$, because of zeros-poles symmetry. Also $z=r e^{i\theta}$
\begin{eqnarray}
u=\ln{\frac{2n r^{n-1}}{1-r^{2n}}}.\nonumber
\end{eqnarray}

\textbf{Example 3.} For $B(z)=\frac{1}{2}( z + \frac{1}{z})$ - the
Joukowsky transform, solution is
\begin{eqnarray}
u(x,y) = \frac{1}{2} \ln \frac{1- 2 (x^2 - y^2) + (x^2 + y^2)^2}{((1-(x^2+y^2))^2-4y^2)^2}.\nonumber
\end{eqnarray}
It is shown in Figure 3.6.

\begin{center}\vspace{0.25cm}
\includegraphics[width=0.50\linewidth]{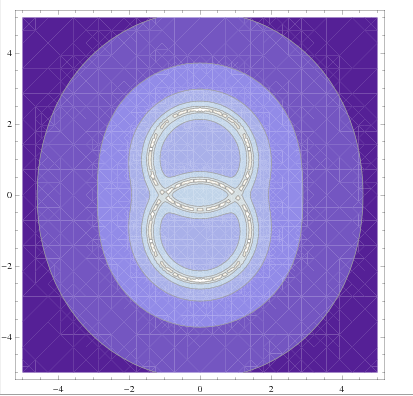}
\captionof{figure}{\color{Navy}Solution of Joukowsky transform}
\end{center}\vspace{0.25cm}

\textbf{Example 4.} Let $B(z)$ is a polynomial function
with $n$ distinct zeros
\begin{eqnarray}
B(z)=(z-z_{1})(z-z_{2})...(z-z_{n}).\nonumber
\end{eqnarray}
Then, using logarithmic function
\begin{eqnarray}
\ln (B(z))=\ln((z-z_{1})(z-z_{2})...(z-z_{n})).\nonumber
\end{eqnarray}
we get
\begin{eqnarray}
\ln (B(z))=\sum_{k=1}^n \ln((z-z_{k})).\nonumber
\end{eqnarray}
To find Liouville solution we need derivatives
\begin{eqnarray}
\frac{1}{B(z)}\frac{d B}{d z}=\sum_{k=1}^n \frac{1}{z-z_{k}}.\nonumber
\end{eqnarray}
This implies
\begin{eqnarray}
\frac{d B}{d z}=B(z) \sum_{k=1}^n \frac{1}{z-z_{k}}.\nonumber
\end{eqnarray}
\begin{eqnarray}
\frac{d \bar B}{d \bar z}= \overline {B(z)} \overline{ (\sum_{k=1}^n \frac{1}{{z-z_{k}}})}.\nonumber
\end{eqnarray}
\begin{eqnarray}
4 \frac{d B}{d z} \frac{d \bar B}{d \bar z}= 4|B(z)|^2  \left|\sum_{k=1}^n\frac{1}{z-z_{k}}\right|^2.\nonumber
\end{eqnarray}
For the  denominator we have
\begin{eqnarray}
(1-B(z) \overline {B(z)})^2= \left(1- \left|\prod_{k=1}^nz-z_{k}\right|^2\right)^2.\nonumber
\end{eqnarray}
Then, combining all this together we have solution of Liouville equation
\begin{eqnarray}
u= \ln \left(\frac{2\prod_{k=1}^n\left|z-z_{k}\right| \left|\sum_{k=1}^n\frac{1}{z-z_{k} }\right|}{1-\prod_{k=1}^n\left|z-z_{k}\right|^2 }\right).\nonumber
\end{eqnarray}
\\
\textbf{Example 4.} Let $B(z)= \sin z$, then applying formula (\ref{**}) we get solution, periodic in $x$:
\begin{eqnarray}
u= \frac{1}{2} \ln \left(\frac{4(\cos^2x+\sinh^2y)}{(\cos^2x+\sinh^2y)^2}\right)  .\nonumber
\end{eqnarray}

\chapter{Conformal Metrics}

\section{Conformal Metrics}
In differential geometry \cite{MG} if we can choose special coordinates
for metric of a surface, such that
\begin{eqnarray}
dl^2 = g(x,y)(dx^2+dy^2),\nonumber
\end{eqnarray}
then these coordinates are called the conformal coordinates. In terms of complex variable $z=x+iy$
we get the metric
\begin{eqnarray}
dl^2 = g(z,\bar z) dz d \bar z.\nonumber
\end{eqnarray}
If $w=w(z)$ (or $z=z(w)$) is a complex analytic coordinate changes, so that,
\begin{eqnarray}
dz = \frac{dz}{d w} d w ,\nonumber
\end{eqnarray}
\begin{eqnarray}
d\bar z = \frac{d\bar z}{d\bar w} d\bar w ,\nonumber
\end{eqnarray}
then the metric remains conformal. Indeed
\begin{eqnarray}
dl^2 = g(z,\bar z)\, \frac{dz}{d w}d w \, \frac{d\bar z}{d\bar w}d\bar w ,\nonumber
\end{eqnarray}
and in terms of
\begin{eqnarray}
G(w, \bar w) = g(z,\bar z)\left|\frac{d z}{d w}\right|^2  ,\nonumber
\end{eqnarray}
we get the conformal metric
\begin{eqnarray}
dl^2 = G(w, \bar w)dwd\bar w .\nonumber
\end{eqnarray}
Below we show two important examples of conformal metrics.
\subsection{Sphere in Conformal Form}
By using stereographic projection of sphere $x_1^2+x_2^2+x_3^2 = R^2$ to complex plane $z$:
\begin{eqnarray}
x_1+ix_2 = R^2 \frac{2z}{R^2+|z|^2},\nonumber
\end{eqnarray}
\begin{eqnarray}
x_3 = R \frac{R^2-|z|^2}{R^2+|z|^2},\nonumber
\end{eqnarray}
we find metric of the sphere $dl^2=dx_1^2+dx_2^2+dx_3^2 $ in conformal form as
\begin{eqnarray}
dl^2= \frac{4 R^4 dz d\bar z}{(R^2+|z|^2)^2}
    = \frac{4 dz d\bar z}{(1+\frac{|z|^2}{R^2})^2}, \label{B}
\end{eqnarray}
where $z=x+iy$.
\subsection{Pseudosphere in Conformal Form}
By using stereographic projection of pseudosphere $-x_1^2-x_2^2+x_3^2 = R^2$ to complex plane $z$:
\begin{eqnarray}
x_1+ix_2 = R^2 \frac{2z}{R^2-|z|^2},\nonumber
\end{eqnarray}
\begin{eqnarray}
x_3 = R \frac{R^2+|z|^2}{R^2-|z|^2},\nonumber
\end{eqnarray}
we find metric of the pseudosphere  $dl^2=-dx_1^2-dx_2^2+dx_3^2 $ in conformal form as
\begin{eqnarray}
-dl^2= \frac{4 R^4 dz d\bar z}{(R^2-|z|^2)^2} = \frac{4 dz d\bar z}{(1-\frac{|z|^2}{R^2})^2}, \label{C}
\end{eqnarray}
where $z=x+iy$.

\section{Metric as Solution of Liouville Equation}

The Gaussian curvature of a surface in conformal form is given by following theorem \cite{MG}.

\textbf{Theorem 1.}  Suppose $(x,y)$ are conformal coordinates with metric
\begin{eqnarray}
dl^2 = g(x,y)(dx^2+dy^2),\nonumber
\end{eqnarray}
then the Gaussian Curvature of surface is
\begin{eqnarray}
K=-\frac{1}{2g(x,y)} \Delta \ln g(x,y).\label{Gaus cu}
\end{eqnarray}

\subsection{Constant Gaussian Curvature Metrics}
If Gaussian Curvature  $K$ of a surface would be constant, then from (\ref{Gaus cu}) follows that metric function $g(x,y)=e^{2u}$ satisfies the Liouville equation $$\Delta u= -K e^{2u}.$$
 The general solution of this equation is
\begin{eqnarray}
u(z, \bar z)=\frac{1}{2}\ln\frac{4\frac{d B}{d z}\frac{d
\bar B}{d \bar z}}{(1+K B(z) \overline {B(z)})^2},\nonumber
\end{eqnarray}
where $B(z)$ is an arbitrary analytic function.
Since  $g(x,y)= e^{2u}$ we get
\begin{eqnarray}
g(z,\bar z)= \frac{4\frac{d B}{d z}\frac{d
\bar B}{d \bar z}}{(1+K B(z) \overline {B(z)})^2},\nonumber
\end{eqnarray}
where $g(x,y)\Rightarrow g(z, \bar z)$. Since
\begin{eqnarray}
dl^2 = g(x,y)(dx^2+dy^2)\nonumber
\end{eqnarray}
\hskip6.8cm\ $ \Downarrow $
\begin{eqnarray}
dl^2 = g(z,\bar z)dz d\bar z,\nonumber
\end{eqnarray}
then the metric is
\begin{eqnarray}
dl^2 = \frac{4\frac{d B}{d z}\frac{d
\bar B}{d \bar z}}{(1+K |B(z)|^2)^2}dzd\bar z. \nonumber
\end{eqnarray}
In terms of $w= B(z)$ it becomes
\begin{eqnarray}
dl^2 = \frac{4 d w d \bar w}{(1+K |w|^2)^2}.\label{A}
\end{eqnarray}\\

Depending on sign of $K$ we have three cases:
\begin{enumerate}
  \item $\textbf{K $>$ 0}$ \emph{Sphere}\\\\
  The metric (\ref{A}) is the metric on sphere in stereographic projection (\ref{B}) with
\begin{eqnarray}
K = \frac{1}{R^2}.\nonumber
\end{eqnarray}

  \item $\textbf{K=0}$  \emph{Euclidean Plane}\\\\
  Euclidean metric on plane is
\begin{eqnarray}
dl^2 = 4\,dw \overline{dw}. \nonumber
\end{eqnarray}

  \item $\textbf{K $<$ 0}$ \emph{Pseudo Sphere}
\end{enumerate}
The metric (\ref{A}) is the metric on pseudosphere in stereographic projection (\ref{C}) with
\begin{eqnarray}
K = -\frac{1}{R^2}.\nonumber
\end{eqnarray}
Choosing $K=-1$, where $g(x,y)=e^{2u}$ satisfies the Liouville equation $\Delta u= e^{2u}$,
then we get the metric
\begin{eqnarray}
dl^2= \frac{4\frac{d B}{d z}\frac{d
\bar B}{d \bar z}}{(1- |B(z)|^2)^2} dzd\bar z
     =\frac{4dBd\bar B}{(1-|B|^2)^2}
     =\frac{4dwd\bar w}{(1-|w|^2)^2},\nonumber
\end{eqnarray}
where $w=B(z)$.
This is the metric for the Poincare model of Lobachevski geometry in unit disk $|w|<1$. Then, solution (\ref{*}) obtained by M\"{o}bius transformation (\ref{mob})
corresponds to Klein model of Lobachevski geometry with metric
\begin{eqnarray}
dl^2= \frac{4\frac{d A}{d z}\frac{d
\bar A}{d \bar z}}{(A(z)+\overline{A(z)})^2} dzd\bar z
     =\frac{4dZd\bar Z}{(Z+\bar Z)^2}\nonumber
\end{eqnarray}
where $Z=A(z)$.
\\
\textbf{Example 1.} Let $B(z)=z$, then we get the metric
\begin{eqnarray}
dl^2= \frac{4}{(1-|z|^2)^2} dzd\bar z .\nonumber
\end{eqnarray}
For cartesian coordinates $z=x+iy$
\begin{eqnarray}
dl^2= \frac{4}{(1-x^2-y^2)^2} (dx^2+dy^2).\nonumber
\end{eqnarray}
This solution describes pseudosphere with $K=-1$ in stereoraphic projection coordinates.

\textbf{Example 2.} For more general choice $B(z)=z^n$, then we have
\begin{eqnarray}
dl^2= \frac{4n^2|z|^{2n-2}}{(1-|z|^{2n})^2} dzd\bar z .\nonumber
\end{eqnarray}
In cartesian coordinates $z=x+iy$ it gives
\begin{eqnarray}
dl^2= \frac{4n^2(x^2+y^2)^{n-1}}{(1-(x^2+y^2)^n)^2} (dx^2+dy^2).\nonumber
\end{eqnarray}
The metric is nonsingular inside the open unit disk $x^2 + y^2 <1$ (the Poincare model).


\section{The Riemann Pseudosphere}
In complex analysis, the common approach to define extended complex plane with unique point at infinity  is the Riemann sphere. Stereographic projection from south (north) pole of the sphere, identifies extended complex plane with the Riemann sphere, where the south (north) pole corresponds to the point at infinity. In a similar way here we introduce another form of extended complex plane in terms of pseudosphere, which we called the Riemann pseudosphere. By stereographic projection, we identify one of the poles of this pseudosphere  with infinity.

\begin{center}\vspace{0.25cm}
\includegraphics[width=0.60\linewidth]{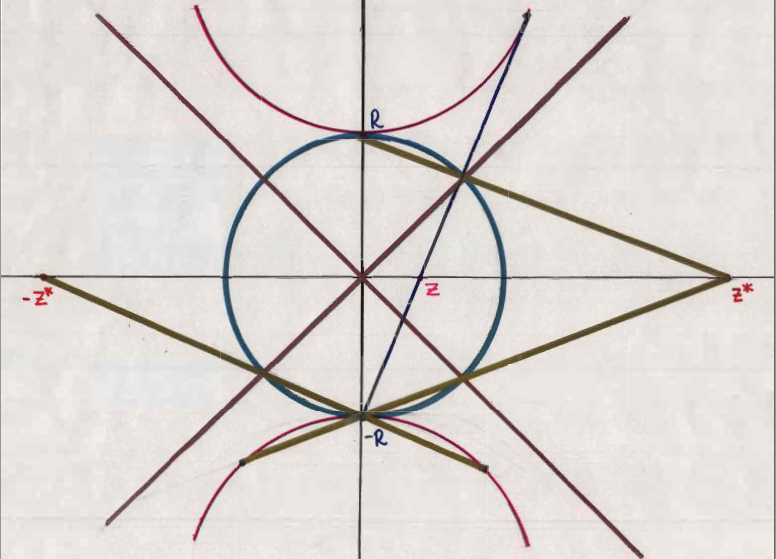}
\captionof{figure}{\color{Navy}The Riemann Sphere and the Riemann Pseudo-Sphere }
\end{center}\vspace{0.25cm}

Equation of pseudosphere $-x_1^2-x_2^2+x_3^2=R^2$ is parameterized by
\begin{eqnarray}
x_1&=& R \sinh\chi \cos\varphi \nonumber \\
x_2&=& R \sinh\chi \sin\varphi \nonumber\\
x_3&=& \pm R \cosh\chi \nonumber
\end{eqnarray}
where $\pm$ sign correspond to the upper and lower parts of pseudosphere correspondingly, also
$0<\chi < \infty$,
$0<\varphi < 2\pi$,
$0< R <\infty$.

By stereographic projection
\begin{eqnarray}
x_1+ix_2 &=& R^2 \frac{2z}{R^2-|z|^2},\nonumber \\
x_3 &=& R \,\frac{R^2+|z|^2}{R^2-|z|^2},\nonumber
\end{eqnarray}
every point in open disk $|z|< R$ is projected to a point in an upper half part of pseudosphere $x_3>0$. And every point, exterior to the disk $|z|>R$ is projected to a point in  lower half part of pseudosphere $x_3<0$. According to this, every point in complex plane, when $|z|\rightarrow \infty$, is going to the south pole $\mathcal{S}$ of the pseudosphere (Figure 4.1). So that the  $\mathcal{S}$ can be considered as a point at infinity for complex plane $\mathbb{C_R}$. We notice that  points at circle $|z|=R$ has no images on pseudosphere. This is why we will consider complex plane $\mathbb{C_R}$ with excluded circle $C: |z|=R $. Such complex plane $\mathbb{C_R}\bigcup \infty$ we called extended complex plane.
Comparing with the Riemann sphere, our Riemann pseudosphere has disadvantages, such that it is not compact and it has no image of the circle with radius $R$. But in some applications, as the Poincare hyperbolic plane model, it could be useful representation. Moreover, due to arbitrariness of radius $R$ we can always choose this radius as an arbitrary small to treat problem of the limit.


\section{Symmetric and Antipodal Points}

\subsection{Symmetric Points on Riemann Sphere} Points
$M(x_1,x_2,x_3)$ and $M'(x_1,x_2,-x_3)$ projected to complex plane
give symmetric points
\begin{eqnarray}
z\,\,\, and\,\,\,z^*=\frac{R^2}{\bar z}\nonumber
\end{eqnarray}
correspondingly. We propose here another way to get symmetric points as projections of the same point $M(x_1,x_2,x_3)$, but  from two different poles. Projection from the south pole $\mathcal{S}(0,0,-1)$ is giving $z$ and from the north pole $\mathcal{N}(0,0,1)$, it is giving $z^*=\frac{R^2}{\bar z}$. This new interpretation gives very simple way of relating symmetric points $z$ and $z^*$ with only one point $M$ on the Riemann sphere (Figure 4.2a).

\subsection{Antipodal Points on Riemann Sphere} Complex numbers $z$
and $-z^*=\frac{R^2}{\bar z}$ are called \textit{antipodal} since
they correspond to projection of points $M(x_1,x_2,x_3)$ and
$M"(-x_1,-x_2,-x_3)$, respectively. These points are opposite points
of diameter of the sphere and antipodals of the Riemann sphere
(Figure 4.2a).

\begin{center}\vspace{0.25cm}
\includegraphics[width=0.80\linewidth]{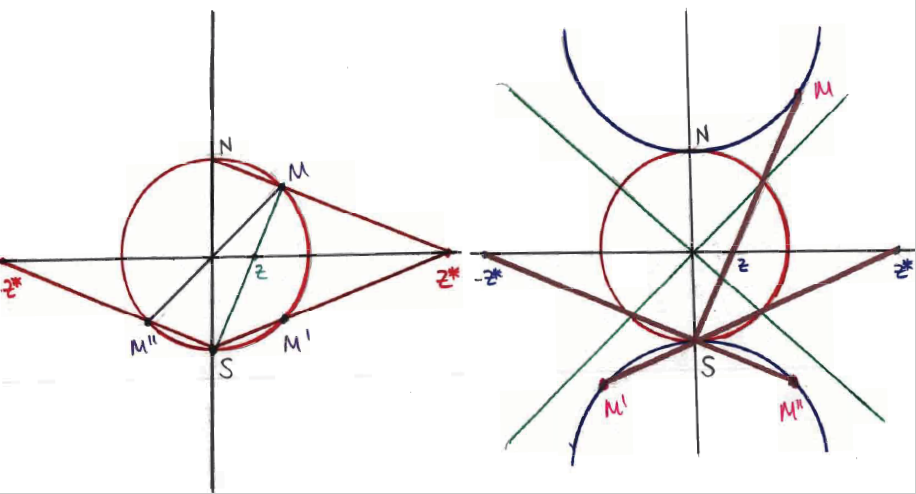}
\captionof{figure}{\color{Navy}Antipodal and Symmetric Points on a) Riemann Sphere; b) Riemann Pseudo-Sphere}
\end{center}\vspace{0.25cm}

\subsection{Antipodal Points on Riemann Pseudosphere} Antipodal
points $z$ and $-z^*=\frac{R^2}{\bar z}$ in complex plane projected
on Riemann pseudosphere, correspond to points $M(x_1,x_2,x_3)$ and
$M"(x_1,x_2,-x_3)$, respectively (Figure 4.2b). This also is in
contrast with the Riemann sphere case.

\subsection{Symmetric Points on Riemann Pseudosphere} Symmetric
points $z$ and $z^*=\frac{R^2}{\bar z}$ in complex plane projected
on Riemann pseudosphere correspond to points $M(x_1,x_2,x_3)$ and
$M'(-x_1,-x_2,-x_3)$, respectively (Figure 3.2b). These points
belong to a line passing from origin and are located at the same
distance from opposite sides. This situation is in contrast with the
Riemann sphere case.

\section{Symmetric Points and Conformal Metric} As we have seen,
conformal metrics for sphere and pseudosphere are
\begin{eqnarray}
dl^2 = \frac{4 dz d \bar z}{(1+K |z|^2)^2},\,\,\,\ K=\pm{\frac{1}{R^2}} ,\nonumber
\end{eqnarray}
respectively.
These metrics are invariant under reflections in $x$ and $y$ axes
\begin{itemize}
  \item $x+iy= z\Rightarrow -z=-x-iy$
  \item $x+iy= z\Rightarrow \,\,\,\bar z =x-iy$
  \item $x+iy= z\Rightarrow -\bar z =-x+iy.$
\end{itemize}
In addition, as easy to check, it is invariant under inversion of symmetric points $z\Rightarrow z^* = \frac{R^2}{\bar z}$. Combining these facts together, we find that it is invariant also under antipodal transformation $z\Rightarrow z^* = -\frac{R^2}{\bar z}$.

\chapter{About So Called The Most General Solution Of Liouville Equation}
Quite recently in paper of D. G. Crowdy \cite{C}, the so called
most general solution of Liouville Equation was presented. Here
we are going to show that this solution can be easily obtained
from the general solution, which is known
from $19^{th}$ century due to Liouville \cite{L}.

\section{D. G. Crowdy's Solution}

Crowdy consider the Liouville equation in the form
\begin{eqnarray}
\psi_{z\bar z} = ce^{d \psi},\label{cdL}
\end{eqnarray}
where $c$ and $d$ are real constants. And he found, as he claim, the most general solution of this equation as
\begin{eqnarray}
\psi=-\frac{2}{d} \log{[c_1Y_1(z)\overline Y_1(\bar z)+c_4Y_2(z)\overline Y_2(\bar z)+c_2Y_1(z)\overline Y_2(\bar z)+\bar c_2 \overline Y_1(\bar z)Y_2(z)]}
 \nonumber \\+ \frac{1}{d} \log{[W(z) \overline W(\bar z)]}\nonumber
\end{eqnarray}
or shortly
\begin{eqnarray}
\psi = \frac{1}{d}\ln{\left(\frac{W \overline W}{ (c_1 Y_1\overline Y_1+ c_4 Y_2\overline Y_2 + c_2\overline Y_2 Y_1 + \bar c_2 Y_2\overline Y_1 )^2}\right)},\label{crowdy}
\end{eqnarray}
where $c_2$ is complex constant, $c_1$ and $c_4$ are real constants.
Here $W = Y_1' Y_2- Y_2' Y_1 $ and relationship between real and complex constants is $cd = -2(c_1c_4-|c_2|^2)$.

Here we like to notice that this solution in fact is determined just by one analytic function. Indeed if $Y_1(z)= H(z) X_1(z)$, $Y_2(z)= H(z) X_2(z)$, then as easy to check $$ W = Y_1' Y_2- Y_2' Y_1 = H^2(X_1' X_2- X_2' X_1)$$ and above solution would be in the same form with replaced $Y$ to $X$. This means that this solution is determined by one function $$B(z) = \frac{Y_1(z)}{Y_2(z)} =\frac{X_1(z)}{X_2(z)}. $$ In the next section we show that it is the M\"{o}bius transformation of the known Liouville solution.

\section{M\"{o}bius Transformation and General Solution of
Liouville Equation}

In section (\ref{231}) we have seen how  a specific M\"{o}bius transformation from half plane (Klein model) to unit disk (Poincare model) connects
two different
solutions of Liouville equation. Now we are going to generalize this
result by considering general M\"{o}bius transformation
\begin{eqnarray}
w = \frac{az+b}{cz+d} ,\,\,\ ad-bc\neq 0.  \nonumber
\end{eqnarray}
It means that we consider relation between analytic functions $A(z)$ and $B(z)$ in the M\"obius form
\begin{eqnarray}
A(z) = \frac{aB(z)+b}{cB(z)+d} ,\,\,\ ad-bc\neq 0.  \label{MBT}
\end{eqnarray}
Substituting $A(z)$ into
\begin{eqnarray}
u=\frac{1}{2}\ln\frac{4\frac{d A}{d z}\frac{d
\bar A}{d \bar z}}{(A(z)+ \bar A(\bar z))^2} \nonumber
\end{eqnarray}
and after simple calculations we get the form \\
\begin{eqnarray}
u = \frac{1}{2}\ln{\left(\frac{4|ad-bc|^2 |B'|^2}{(|B|^2(a\bar c+\bar a c)+\bar B(d\bar a + b\bar c)+B(\bar d a+\bar b c)+(\bar d b+ d \bar b))^2}\right)}.  \nonumber
\end{eqnarray}
From Complex Analysis  we know \cite{A} that any M\"{o}bius
transformation
\begin{equation}
w = \frac{a z + b}{c z + d}
\end{equation}
can be related with linear transformation
\begin{eqnarray}
\left( \begin{array}{c} w_1 \\
                        w_2     \\
                        \end{array}\right)  =
\left( \begin{array}{cc} a & b  \\
                         c & d    \\
                         \end{array}\right)
                                 \left( \begin{array}{c} z_1 \\
                                                         z_2     \\
                                 \end{array}\right) ,\label{lt1}
\end{eqnarray}
by substitution $w=\frac{w_1}{w_2}$ and $z=\frac{z_1}{z_2}$. Coordinates $z_1$, $z_2$ are called the homogeneous coordinates.

According to this observation, the linear transformation
\begin{eqnarray}
\left( \begin{array}{c} X_1 \\
                        X_2     \\
                        \end{array}\right)  =
\left( \begin{array}{cc} a & b  \\
                         c & d    \\
                         \end{array}\right)
                                 \left( \begin{array}{c} Y_1 \\
                                                         Y_2     \\
                                 \end{array}\right) \label{lt1}
\end{eqnarray}
of homogeneous coordinates $X_1(z), X_2(z)$ and $Y_1(z), Y_2(z)$ as an arbitrary analytic functions,
in terms of
\begin{eqnarray}
A(z) = \frac{X_1(z)}{X_2(z)},  \nonumber
\end{eqnarray}
\begin{eqnarray}
B(z) = \frac{Y_1(z)}{Y_2(z)},  \nonumber
\end{eqnarray}
implies the  M\"{o}bius transformation (\ref{MBT}).

The Meromorphic form of solution $A(z)= \frac{C(z)}{D(z)}$, which we have discussed in section (\ref{221}), is also in this form.
Considering $B(z)$ as a ratio of two analytic functions
\begin{eqnarray}
B(z) = \frac{Y_1(z)}{Y_2(z)}  \nonumber
\end{eqnarray}
and denoting $\alpha_1=(a\bar c+\bar a c)$, $\alpha_2=(\bar d a+\bar b c)$, $\alpha_4=(\bar d b+ d \bar b),$
after  substitution we get solution in the form \\
\begin{eqnarray}
u = \frac{1}{2}\ln{\left(\frac{4|ad-bc|^2 |Y_1' Y_2- Y_2' Y_1|^2}{(|Y_1|^2 \alpha_1 +\overline Y_2 Y_1 \alpha_2 + Y_2\overline Y_1\bar \alpha_2 + |Y_2|^2 \alpha_4)^2}\right)}.  \label{Y1}
\end{eqnarray}
To compare this solution with the one given by Crowdy (\ref{crowdy}), we identify  $W = Y_1' Y_2- Y_2' Y_1 $ and
\begin{eqnarray}
c_1=\frac{\alpha_1}{2|ad-bc|},\,\,
c_2=\frac{\alpha_2}{2|ad-bc|},\,\,
c_4=\frac{\alpha_4}{2|ad-bc|}. \nonumber
\end{eqnarray}
Then solution (\ref{Y1}) becomes \\
\begin{eqnarray}
u = \frac{1}{2}\ln{\left(\frac{|W|^2}{ (c_1 |Y_1|^2 + c_4|Y_2|^2 + c_2\overline Y_2 Y_1 + \bar c_2 Y_2\overline Y_1 )^2}\right)}.  \nonumber
\end{eqnarray}
This solution coincides with the one given by Crowdy,  where $c=\frac{1}{4}$ and $d=2$. (Generalization to arbitrary $d$ is straightforward as was shown in Section 2.2.2). As easy to see, condition $cd = -2(c_1c_4-|c_2|^2)$ is also satisfied. This shows that solution of Crowdy, as the most general solution of Liouville equation, is just M\"{o}bius transformation in homogeneous coordinates of the general solution obtained by Liouville.

\section{Most General Solution for Constant Gaussian Metric}

For surfaces with constant Gaussian curvature $K$,  the
Liouville equation is \\ $\Delta u = -Ke^{2u}$ or
$$u_{z\bar
z}=-\frac{K}{4}e^{2u}. $$
It is in the form (\ref{cdL}), where $c=-\frac{K}{4}$, $d=2$. The general
solution of this equation is
\begin{eqnarray}
u(z,\bar z) = \frac{1}{2}\ln{\left(\frac{4|B'(z)|^2}{ (1+K|B(z)|^2)^2}\right)}.  \nonumber
\end{eqnarray}
Applying M\"{o}bious transformation
\begin{eqnarray}
B(z) = \frac{p\,\, Y(z)+r}{s\,\, Y(z)+m} ,\,\,\ pm-rs\neq 0.  \nonumber
\end{eqnarray}
after simple calculations we get
\begin{eqnarray}
u = \frac{1}{2}\ln{\left(\frac{4|pm-sr|^2 |Y'|^2}{(|Y|^2(|s|^2+K|p|^2)+Y(s\bar m+ K p\bar r)+\bar Y(\bar s m+ K\bar p r) + (|m|^2+K |r|^2))^2}\right)}.  \nonumber
\end{eqnarray}

We denote
\begin{eqnarray}
Y(z) = \frac{T_1(z)}{T_2(z)}  \nonumber
\end{eqnarray}
and $\alpha_1= |s|^2+K |p|^2$, $\alpha_2=(s \bar m+ K p\bar r)$, $\alpha_4=(|m|^2+ K |r|^2)$.
Then by substitution we find solution in the form \\
\begin{eqnarray}
u = \frac{1}{2}\ln{\left(\frac{4|pm-sr|^2 |T_1' T_2- T_2' T_1|^2}{(|T_1|^2 \alpha_1 +\overline T_2 T_1 \alpha_2 + T_2\overline T_1\bar \alpha_2 + |T_2|^2 \alpha_4)^2}\right)}.  \label{T1}
\end{eqnarray}
To compare this solution with the one given by Crowdy (\ref{crowdy}), we identify  $W = Y_1' Y_2- Y_2' Y_1 $ and
\begin{eqnarray}
c_1=\frac{\alpha_1}{2|pm-sr|},\,\,
c_2=\frac{\alpha_2}{2|pm-sr|},\,\,
c_4=\frac{\alpha_4}{2|pm-sr|}. \nonumber
\end{eqnarray}
Then solution (\ref{T1}) becomes \\
\begin{eqnarray}
u = \frac{1}{2}\ln{\left(\frac{|W|^2}{ (c_1 |T_1|^2 + c_4|T_2|^2 + c_2\overline T_2 T_1 + \bar c_2 T_2\overline T_1 )^2}\right)}.  \nonumber
\end{eqnarray}
This coincides with Crowdy,  where $c=\frac{-K}{4}$ and $d=2$. As easy to see, condition $cd = -2(c_1c_4-|c_2|^2)$ is satisfied as well.
\chapter{Conclusions}
In present paper we have introduced the  Nonlinear Cauchy Riemann
equations and corresponding linear and nonlinear equations of
Laplace type. By using it, we constructed general solution of Liouville
equation and found relation of it with conformal metric on surfaces with constant Gaussian curvature.
By M\"{o}bius transformation, relation between several geometries and
most general solution of Liouville equation were derived.

We note that proposed here
Nonlinear Cauchy Riemann equations can be applied to study solutions of
other linear and nonlinear equations like the Helmholtz equation and
the Sine-Gordon equation. Moreover, physical interpretation of these
equations in terms of hydrodynamic flow can be studied. These questions are under investigation.

\section{Acknowledgements} This work is supported by TUBITAK grant 116F206.

\end{document}